\documentstyle[twocolumn,prl,floats,aps,epsfig]{revtex}
\begin{document}

\draft

\wideabs{
\title{Observation of a Universial Low Temperature Limit of the Free
Flux Flow Resistivity in Superclean Bi$_2$Sr$_2$CaCu$_2$O$_8$
Whiskers}

\author{Weimin Chen, J.P. Franck, and J. Jung}

\address{Department of Physics, University of Alberta, Edmonton, Canada T6G 2J1}

\date{\today}

\maketitle

\begin{abstract}
We report peculiar vortex motion in clean
Bi$_2$Sr$_2$CaCu$_2$O$_8$ single crystals, which reveals nearly
linear-$T$ resistivity tails extending to low temperatures. The
resistivity was found to satisfy the classical Bardeen-Stephen
model for free flux flow, which is otherwise hard to observe under
normal experimental conditions in high-$T_c$ cuprates. Moreover,
in the superclean regime, a universal temperature-independent
value for the flux flow resistivity was observed, $\rho_f =
1.4(B-B^*)$. We interpret this carrier-relaxation {\it
independent} behavior as being related to the dissipation caused
by the minigap states within the vortex core. The results are in
agreement with theoretical calculations (N. B. Kopnin and G. E.
Volovik, Phys. Rev. Lett. {\bf 79}, 1377 (1997)).
\end{abstract}

}

Free flux flow of vortices in the mixed state of superconductors
describes the motion of vortices under the influence of Lorentz
force and an effective viscosity. The vortices move perpendicular
to the applied current and induce an EMF in the current direction.
This leads to a sizable magnetoresistance and a very small Hall
effect \cite{1}. The classical Bardeen-Stephen (BS) theory
\cite{2} describes this phenomenon in terms of the normal-state
resistivity and normal electrons in the vortex core. Caroli, de
Gennes, and Matricon \cite{3} realized that quantizied states with
an energy gap $\omega_0$ exist in the vortex core. In the case of
large impurity scattering, these states are smeared and the BS
theory gives the correct description of free vortex flow.  In the
"superclean" case, as measured by $\omega_0$$\tau$$\gg1$, where
$\tau$ is the normal-state relaxation time, the core levels are
distinct and the vortex motion in a field becomes parallel to the
external current. This leads to an increasing Hall effect and a
diminishing flux flow resistivity for conventional ($s$-wave)
superconductors.  In the case of $d$-wave superconductors, it was
proposed by Kopnin and Volovik \cite{4} that the flux flow
resistivity assumes at low temperature a universal value
independent of the normal-state relaxation time. Nodes in the
superconducting gap of a $d$-wave superconductor lead to node
structure of the vortex gap, resulting in resonance absorption of
zero frequency vortex modes.  An approach to a Hall angle of
$\pi/2$ has been observed \cite{5} in a 60 K YBCO single crystal.
In this work we report the possible observation of the resistivity
limit in very clean single crystalline whiskers of
Bi$_2$Sr$_2$CaCu$_2$O$_8$ (Bi-2212) high-$T_c$ superconductors.

In order to observe free flux flow, one must have samples with
very low concentration of pinning centers.  In high-$T_c$
superconductors, this condition is difficult to obtain due to the
microscopic layered structure, the small coherence length, and the
different order parameter.  Because of this, free flux flow has
been observed only in narrow ranges of field and temperature
\cite{6} or for pulsed large current density \cite{7}.  In
general, thermally activated flux flow (TAFF) is observed
\cite{8}, followed by the formation of a vortex solid at low
temperatures \cite{9}. In this paper we report the properties of
two Bi-2212 whiskers, both showing free flux flow but with
different field dependencies.  The differences in
magnetoresistance can be correlated with the moderately clean and
superclean regimes reached.  In the superclean regime, we observed
at low temperatures the universal flux flow resistivity which is
independent of carrier relaxation \cite{4}.

The whiskers were grown by long-time annealing at temperatures
between 835 and 855$^\circ$C from ceramic (and non-stoichiometric)
pellets of BiSrCaCu oxide as described previously \cite{10,11}.
The first whisker (Whisker 1) described here was from a pellet
grown at atmospheric pressure (20\% O$_2$), and can be assumed to
be overdoped. The resistive $T_c$ is 74.6 K. The second whisker
(Whisker 2) was also initially grown at atmospheric pressure. It
was then annealed in flowing nitrogen at 500$^\circ$ for 4 hours.
This whisker is underdoped with a resistive $T_c$ of 73.2 K, a
pseudogap behavior near 220 K was observed in the normal-state
resistivity in support of this assessment. The geometry of the
whiskers and the attached electrical leads were measured by a
scanning electron microscope. The dimensions are as follows
(length measured between potential leads): Whisker 1: 0.654 mm
$\times$ 4.9 $\mu$m $\times$ 0.48 $\mu$m (thickness); Whisker 2:
0.28 mm $\times$ 4.38 $\mu$m $\times$ 0.20 $\mu$m (thickness). The
three crystallographic axes ($a$, $b$, and $c$) are along these
directions, respectively \cite{11}.

Resistance measurement was done by the standard four-probe method.
Electrical leads were fabricated by sputtering silver stripes onto
the whisker followed by attaching copper wires with indium.  We
used true dc current rather than ac modulation in order to
minimize disturbance to the flux flow.  Each measurement was done
by driving the current in both directions for a few times and
averaging the voltages so as to reduce the thermal noise.  The
current is always along the $a$-axis and fixed at 0.5 $\mu$A
(current density $\sim 10$ A/cm$^2$), whereas the magnetic field
was applied along all the three axes and changed between 0 and 5.5
T. One notices that the sample thickness is of the order or
smaller than the penetration depth (2600 to 3000 \AA\ in Bi-2212).
Because of this, and the large demagnetization effect in the $H
\parallel c$ configuration, we assume that the applied field $H$
equals the magnetic induction $B$ in the whiskers.

Figures 1 and 2 show the resistivities for both whiskers in
different fields up to 5.5 T. Whisker 2 was only measured in
fields of 3, 4, and 5.5 T. We can  see that an extended region,
where $\rho(T,B)$ is linear in $T$, appears below the extensive
fluctuation range near $T_c$ and extends to very low values of
$\rho(T,B)$. No sign of TAFF or flux line melting was observed in
either whisker.  Flux line melting occurs only in fields of 400 G
or below and for $T \gtrsim 40$ K \cite{9,12}, so it is not
surprising that we did not observe it. We interpret the vortex
motion associated with the long linear $\rho-T$ tails as being in
the free flux flow regime. The characteristics of the $\rho(T,B)$
for both whiskers in this regime show certain similarities, as
well as differences.

The free flux flow regime starts at a characteristic low
temperature for a constant field. We call this field $B^*(T)$,
below which the resistance is zero due to immobilization of the
vortices. The temperature dependence of $B^*$ could be fitted to:

\begin{equation}
B^*=B_{0}^{*}(1-T/T_c)^n.
\end{equation}
For Whisker 1, we found $n = 3.33 \pm 0.05$ and $B_{0}^{*} =
24.45$ T. For Whisker 2, the same fitting with fixed $n = 3.33$
yielded a much smaller $B_{0}^{*} = 6.71$ T, although the exponent
is less certain here due to the fewer fields used ($n = 3.3 \pm
0.3$). All the applied fields are well above the vortex melting
line. The fields $B^*(T)$ are close to the depinning line, which
is dependent on the amount of disorder in the samples \cite{12}.
In spite of the very similar zero-field transition temperatures,
we see that the two whiskers exhibit different properties.  In
particular, Whisker 2 shows magnetoresistance down to very low
temperatures, close to 0 K! If the fitting of eq. (1) can be
extrapolated, one would expect non-zero flux flow resistance at 0
K in this whisker at $H=6.7$ T. Extension of the free flux flow
regime to such low temperatures has not previously been observed.

A second difference between the two whiskers is the field
dependence of the magnetoresistivity.   For Whisker 1,
$\rho_f/\rho_n$ is linear in fields higher than about 1.5 T and in
the temperature range from 40 to 60 K.  At lower fields, a
curvature appears in $\rho_f$ due to the 3D to 2D crossover. We
use then the Bardeen-Stephen model \cite{2} in its differential
form in the high-field regime:

\begin{equation}
{d\rho_f/\rho_n \over {dB}} = {1 \over B_{c2}}.
\end{equation}
This corresponds to the field region for pancake vortices with
small interplane interaction. For $H \gtrsim 1.5$ T, we obtained
reasonable $B_{c2}$ values both for $H \parallel c$ (in the range
of 40 to 60K ) and $H \parallel b$ (65 to 70 K). As expected, the
fields extrapolate to $T_c$ with slopes of -1.33 T/K ($H
\parallel c$) and -30.8 T/K ($H
 \parallel b$), as shown in Fig. 3.  These data are in good
agreement with literature values both for the temperature
dependence and the anisotropy \cite{13}.

In the case of Whisker 2 no such analysis is possible.  We show in
Fig. 4 both $\rho_f$ and $\rho_f/\rho_n$ as a function of field
for this whisker. As can be seen, $\rho_f/\rho_n$ and $\rho_n$ at
high temperatures are only weakly dependent on field above 3 T. At
lower temperatures, the dependence of the flux flow resistivity
$\rho_f$ on field increases, and assumes below approximately 30 K
a uniform, temperature independent slope of $d\rho_f/dT = 1.40$
$\mu \Omega$cm/T. Attempts to fit to the Bardeen-Stephen
expression of eq. (2) would lead to unacceptable values of
$B_{c2}$ and wrong temperature dependence.

The characteristic behavior of $\rho_f$ in Whisker 2 is the
near-linearity of $\rho_f$ for fields above 3 T.  Below 3 T we
have extrapolated $\rho_f$ at each temperature to $B^*$ calculated
from eq. (1).  This involves curvature which probably again is
connected to the 3 D to 2 D crossover at high fields.  At
temperatures below about 30 K, $\rho_f(B)$ has the same slope, and
extends to $B^*$ for $T \leq 15$ K.  In this regime we have,
therefore:

\begin{equation}
\rho_f(T)=\rho_0(B-B^*),
\end{equation}
where the constant is given by $\rho_0 = 1.40 \pm 0.1$ $\mu\Omega$
cm/T. Above $T \sim 30$ K, the slope decreases and reaches 0.58
$\mu\Omega$cm/T at 55 K, or one third of the low-temperature
limiting value.

The different magnetoresistive behavior of the two whiskers are
related to their microscopic properties.  We first discuss their
normal-state properties.  For Whisker 1,  a linear temperature
dependence of $\rho_n$ (in zero field) was obtained down to the
fluctuation regime near $T_c$. We base our analysis on an
extrapolation of this linear part to the lowest temperatures. The
axis intercept at 0 K is about 23.0 $\mu\Omega$cm, or about 6.4\%
of the resistivity at 300 K.  The two-dimensional square
resistance $R_\Box$ per CuO$_2$ layer is $R_\Box = \rho_n/d$ with
$d = 15$ \AA\ for Bi-2212. $R_\Box$ varies between 2390 $\Omega$
at 300 K to 190 $\Omega$ at 5 K. From this we calculate the
product $k_Fl = h/e^2R_\Box$, where $k_F$ is the radius of Fermi
surface and $l$ the mean free path. $k_Fl$ obtained in this way
varies between 10.8 at 300 K and 135 at 5K. At low temperatures,
Whisker 1 is therefore in the moderately clean regime. One can
estimate $l$ from this by using a reasonable value for $k_F$
($\approx 0.33$ \AA$^{-1}$), $l$ then reaches a value of 330 \AA\
at 5 K. For Whisker 2, the linear part of $\rho_n$ extrapolates to
a nearly zero $\rho$(0 K) ($\pm 5$ $\mu \Omega$cm). $R_\Box$ in
this case changes from 3400 $\Omega$ (at 300 K) to 56.7 $\Omega$
(at 5 K), and accordingly, $k_Fl$ varies from 7.6 to 456. This
leads to an estimate for $l$ ($k_F = 0.33$ \AA$^{-1}$) of 1110
\AA\ at 5 K. Therefore, Whisker 2 enters further into the clean or
superclean regime at low temperatures, although this occurs only
below 30 K. It is, however, important to notice that in the field
of 5.5 T, we can explore further into the superclean regime with
Whisker 2. A more drastic difference between the two whiskers is
evident in the minigap ($\omega$) of the quasiparticle states
within the vortex core \cite{4,14}. The minigap is given by:

\begin{equation}
\omega_0=\frac{\Delta_{0}^{2}}{\varepsilon_F}
\end{equation}
where $\Delta_0$ is the (maximum) superconducting gap at 0 K, and
$\varepsilon_F$ the Fermi energy.  We use $\varepsilon_F  = 500$
meV for both whiskers \cite{15}. The value of $\Delta_0$ is,
however, drastically different. For Whisker 1 (overdoped), we
estimate $\Delta_0$ = 10 meV, and for Whisker 2 (underdoped)
$\Delta_0$ = 25 meV \cite{16}.  The minigaps are then 0.20 meV for
Whisker 1, and 1.25 meV for Whisker 2.  The region of superclean
behavior is governed by the size of the product $\Gamma = \omega_0
\tau$, where $\tau$ is the normal-state relaxation time.  The
approximately six-fold increase of $\omega_0$ for Whisker 2 over
that of Whisker 1 is remarkable.  We can further estimate $\tau$
from

\begin{equation}
\tau=\frac{4\pi \lambda_{0}{2}}{c^2\rho_n}
\end{equation}
where $\lambda_o$ is the penetration depth at 0 K.  We use
$\lambda_0 = 2600$ \AA\ for Whisker 1 \cite{16}, and a slightly
larger $\lambda_0 = 3000$ \AA\ for Whisker 2 \cite{17}. For
Whisker 1, the relaxation time obtained from eq. (5) reaches 0.37
psec at the lowest temperatures, and $\Gamma = \omega_0 \tau=0.1$.
For Whisker 2, on the other hand, $\tau = 6.64/T$ (psec), so that
$\tau$ reaches very large values at low temperatures (2.54 psec
for $T =5$ K). It appears, therefore, that in this whisker, the
superclean region is reached at low temperatures. All the
estimates are based on the extrapolated normal-state resistivity,
without adjustment for a possible reduction of $\rho_n$ in the
vortex cores.  The parameter $\Gamma$ is, therefore, at low
temperatures considerably larger for Whisker 2 than for Whisker 1.
This is due to the combination of a larger $\omega_0$ and a
smaller $\rho_n$ for Whisker 2. Estimates of $\Gamma$ and $k_Fl$
are shown in Fig. 5.

In the moderately clean regime, the free flux flow ohmic
resistivity is given by Kopnin and Volovik \cite{4} as:

\begin{equation}
\rho_f={B \over n_{h}ec\Gamma\ln(T_c/T)}.
\end{equation}
In the superclean regime the free flux flow resistivity assumes a
universal non-zero limit, independent of the relaxation time (and
hence normal-state resistivity), which is given by:

\begin{equation}
\rho_f = {\pi B \over 2n_hec}.
\end{equation}
The temperature at which the moderately clean value of $\rho_f$
equals the low temperature ideal limit was estimated by using the
above value of $\Gamma$  for Whisker 2, and is found to be $T$ =
23 K. This is in good agreement with our observations.

Since in Whisker 2 at low temperatures, the free flux flow sets in
at $B^*$ and is linear in $B$, we use eq. (7) in a modified form

\begin{equation}
\rho_f={\pi (B-B^*) \over 2n_hec}.
\end{equation}
The only system parameter here is $n_h$, which can be estimated
from the zero-temperature penetration depth $\lambda_0 = 3000$
\AA\ as $n_h = 0.63 \times 10^{21}$ cm$^{-3}$. We would then
expect from eq. (8) a limiting free flux flow resistivity of
$1.56(B-B^*)$ $\mu \Omega$cm. This compares extremely well with
the observed value of $1.4(B-B*)$ $\mu \Omega$cm. The observed
limiting resistivity thus most likely represents the predicted
universal behavior. This result is connected to the gap nodes in a
$d$-wave superconductor, which results in higher order nodes in
the minigap \cite{5}.

There have been several observations of free flux flow in
moderately clean or superclean high-$T_c$ superconductors by
Matsuda {\it et al.} \cite{18}, Harris {\it et al.} \cite{5}, and
Doettinger {\it et al.} \cite{19}.  Harris {\it et al.} studied
the magnetoresistance in an oxygen deficient single crystal of
YBCO with $T_c$ = 64 K in fields up to 24 T. They find that the
normally small Hall angle increases with falling temperatures and
reaches 70$^\circ$ at 13 K. This increase is due to the expected
deviation of vortex motion from perpendicular to the current
towards the current direction with increasingly clean conditions
characterized by $\omega_0\tau$. The same effect leads to the
reduction in the ohmic component of the flux flow resistance.  The
ohmic resistivities of Ref. 5 are indeed very similar to those
seen from Whisker 2. At temperatures below about 30 K and in
fields above 10 T, they observed a slope $d\rho_f/dB$ of 2.4
$\mu\Omega$cm/T. This is similar to what we found, but larger than
the theoretical value of 1.42 $\mu\Omega$cm/T from eq. (8) ($n_h$
is estimated from $\lambda_0 = 2500$ \AA\ \cite{20} for
oxygen-deficient YBCO). The linear dependence of the Hall angle on
$\omega_0\tau$ observed by them suggests that the universal limit
is not quite reached in their sample.

In conclusion, we observed free flux flow behavior at low
temperatures in the magnetoresistance of Bi-2212 whiskers.  For an
underdoped sample, we observed free flux flow down to 5.4 K in a
field of 5.5 T. Below about 30 K, the free flux flow resistivity
assumes a universal, temperature-independent value, in numerical
agreement with the calculations of Kopnin and Volovik \cite{4}.

We gratefully acknowledge support of this work by the National
Sciences and Engineering Research Council of Canada

Figure Captions

Fig. 1. The resistivity of Whisker 1 in fields between 0 and 5.5
T.  Solid curves: $H \parallel c$; dotted curves: $H \parallel b$.

Fig. 2  The resistivity in 3, 4 and 5.5 T of Whisker 2.  Fits to
the linear-$T$ regime are shown.

Fig. 3  Estimates of $B_{c2}$ for Whisker 1 obtained from the
Bardeen-Stephen Theory, eq. (2).

Fig. 4  The resistivity in the free flux region of Whisker 2.  (a)
Resistivity $\rho_f$; (b) Reduced resistivity $\rho_f/\rho_n$,
where $\rho_n$ is the extrapolated normal-state resistivity. The
dotted lines are rough extrapolations to the onset field from eq.
(1).

Fig. 5  Estimates of $k_Fl$ and $\Gamma = \omega_0 \tau$ for
Whiskers 1 and 2. The regimes over which we observed free flux
flow are shown by the heavy segments.


\begin{references}
\bibitem{1}C. F. Hempstead and Y. B. Kim, Phys. Rev. Lett. {\bf 12}, 145 (1964);
Y. B. Kim, C. F. Hempstead, and A. R. Strnad, Phys. Rev. {\bf
131}, 2486 (1963).

\bibitem{2} J. Bardeen and M. J. Stephen, Phys. Rev. {\bf 140}, A1197 (1965).

\bibitem{3} C. Caroli, P. G. de Gennes, and J. Matricon, Phys. Lett. {\bf 9},
307 (1964).

\bibitem{4} N. B. Kopnin and G. E. Volovik, Phys. Rev. Lett. {\bf 79}, 1377 (1997).

\bibitem{5} J. M. Harris, Y. F. Yan, O. K. C. Tsui, Y. Matsuda, and N. P. Ong,
Phys. Rev. Lett. {\bf 73}, 1711 (1994).

\bibitem{6}  J. A. Fendrich {\it et al.}, Phys. Rev. lett. {\bf 74},
1210 (1995).

\bibitem{7} N. B. Kunchur, D. K. Christen, and J. M. Phillips, Phys. Rev. Lett.
{\bf 70}, 998 (1993).

\bibitem{8} T. T. M. Palstra, B. Batlogg, L. F. Schneemeyer, and J. V. Waszczak,
Phys. Rev. Lett. {\bf 61}, 1662 (1988).

\bibitem{9} R. Cubitt {\it et al.}, Nature (London) {\bf 365}, 407 (1993).

\bibitem{10} J. Jung, J. P. Franck, D. F. Mitchell, and H. Claus, Physica C
{\bf 156}, 494 (1988).

\bibitem{11} J. Jung, J. P. Franck, S. C. Cheng, and S. S. Sheinin, Jpn. J. Appl .
Phys. {\bf 28}, L1182 (1989).

\bibitem{12} B. Khaykovich {\it et al.}, Phys. Rev. B {\bf 56},
R517 (1997).

\bibitem{13} Lu Zhang, J. Z. Liu, and R. N. Shelton, Phys. Rev. B {\bf 45}, 4978
 (1992).

\bibitem{14} N. B. Kopnin and A. V. Lopatin, Phys. Rev. B {\bf 51}, 15291 (1995).

\bibitem{15} Z.-X. Shen and D. S. Dessau, Phys. Reports
{\bf 253}, 1 (1995).

\bibitem{16} J. M. Harris, Z.-X. Shen, P. J. White, D. S. Marshall,
and M. C. Schabel, Phys. Rev. B {\bf 54}, R15665 (1996).

\bibitem{17} T. Jacobs, S. Sridhar, Q. Li, G. D. Gu, and N. Koshizuka, Phys. Rev.
Lett. {\bf 75}, 4516 (1995).

\bibitem{18} Y. Matsuda {\it et al.}, Phys. Rev. B {\bf 49}, 4380 (1994).

\bibitem{19} S. G. Doettinger, R. P. Huebener, S. Kittleberger, and C. C. Tsuei,
Europhys. Lett. {\bf 33}, 641 (1996).

\bibitem{20} Y. G. Uemura {\it et al.}, Phys. Rev. B {\bf 38}, 909 (1988).
\end{references}
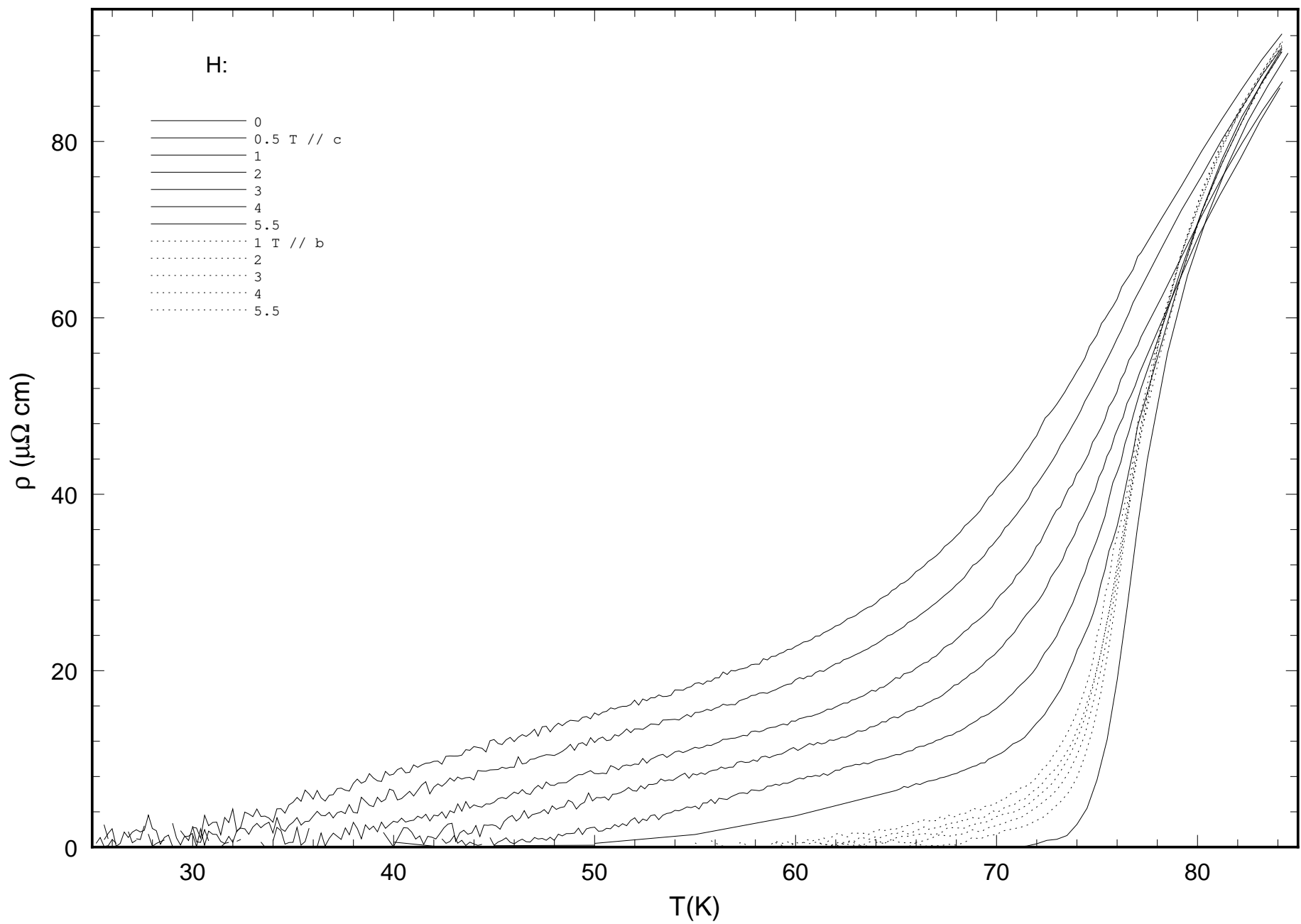
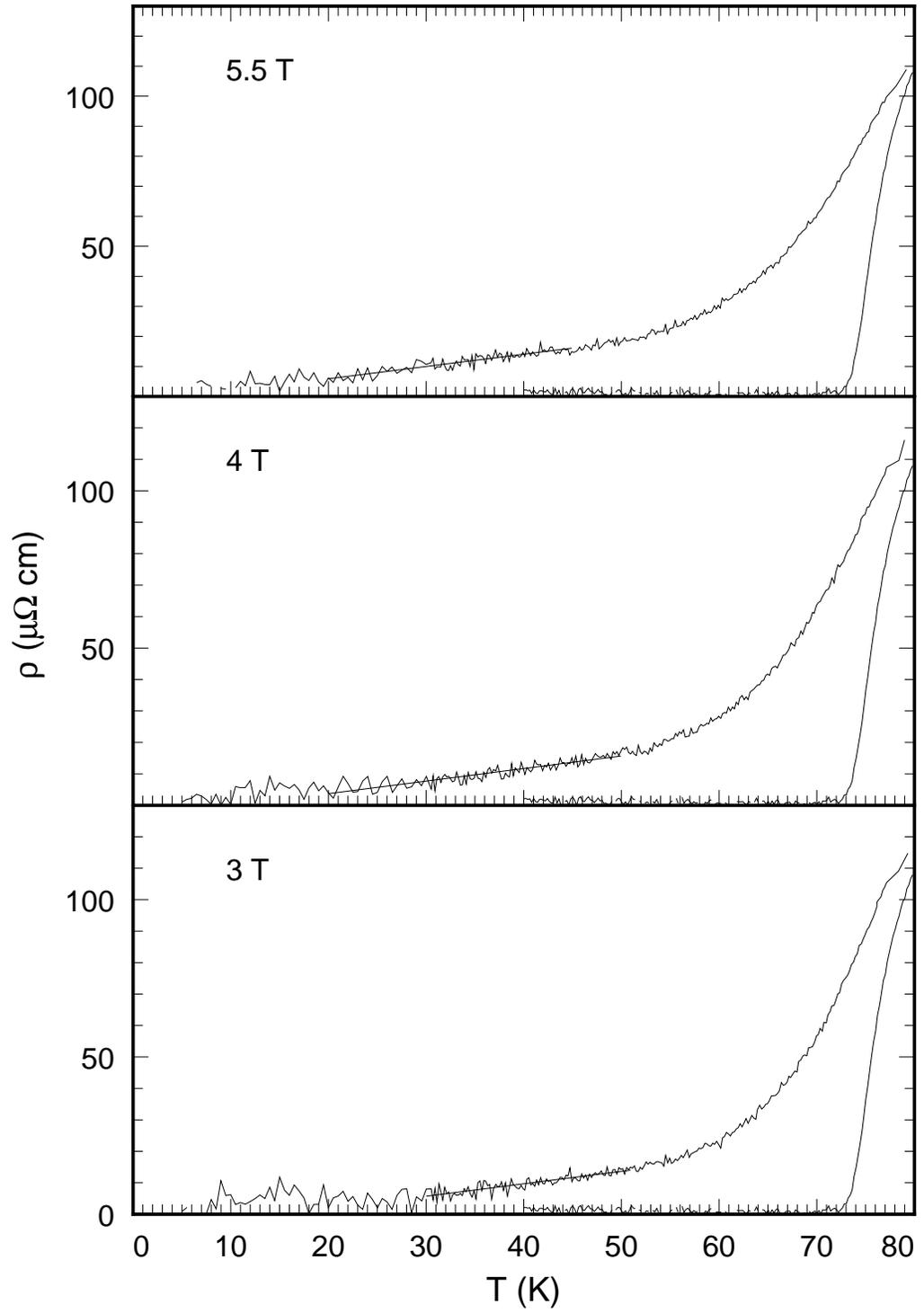
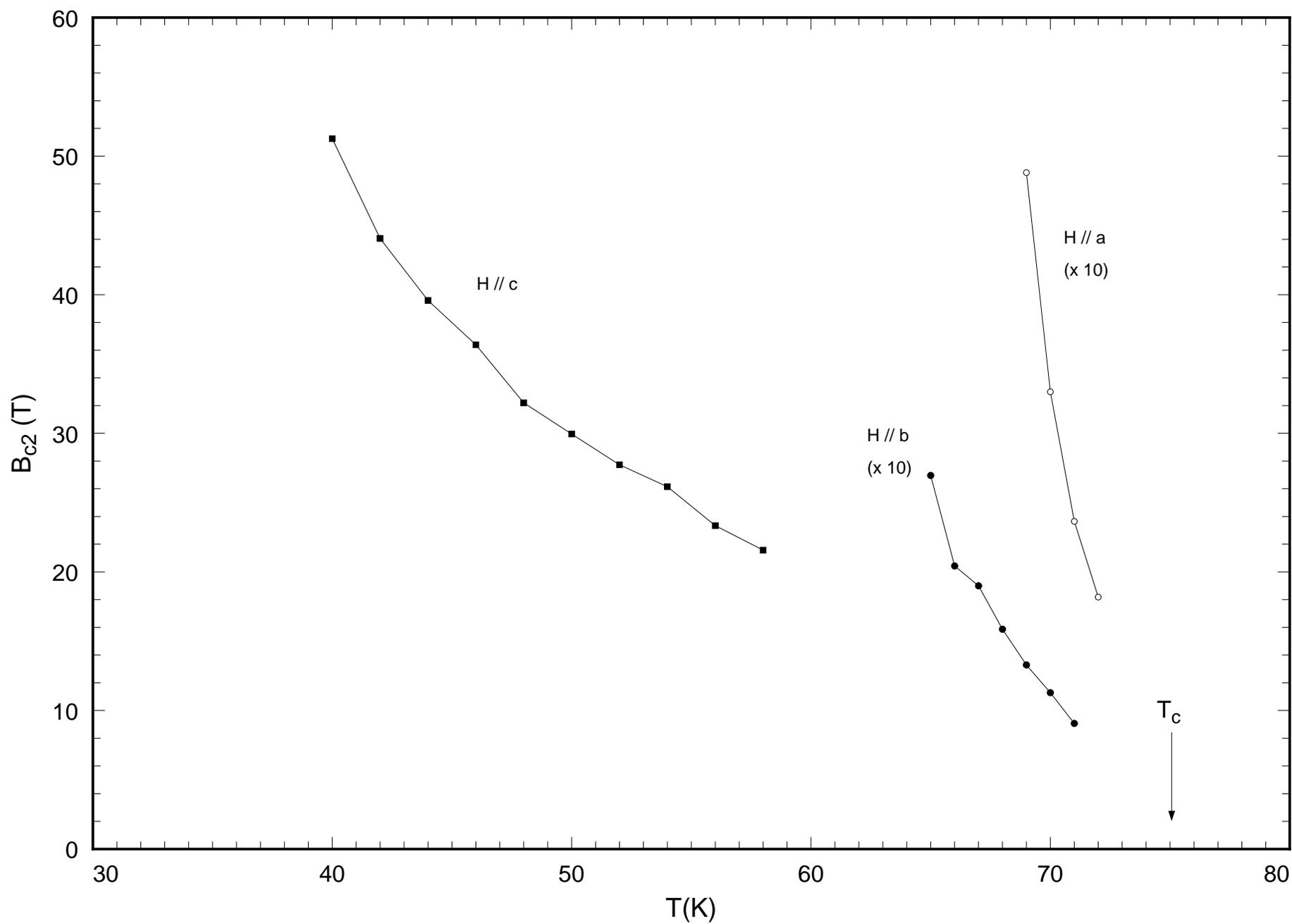
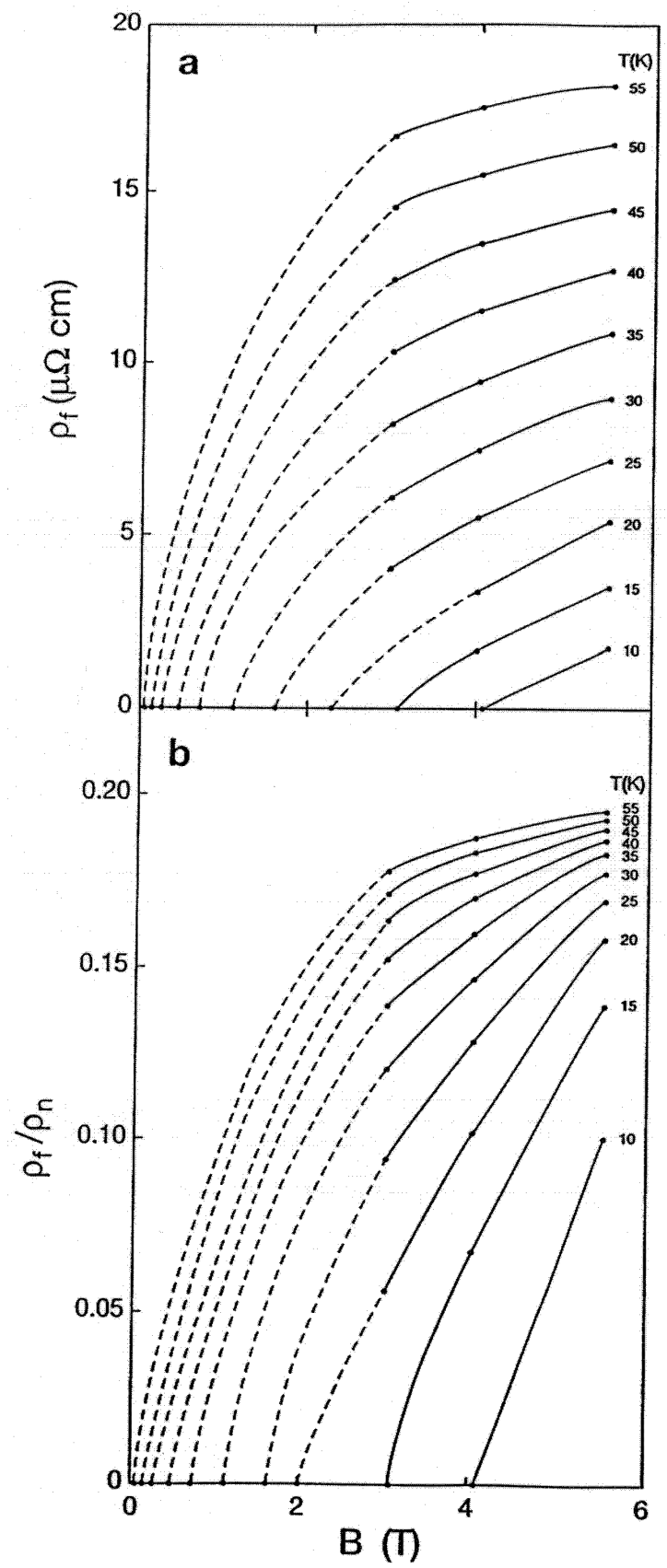
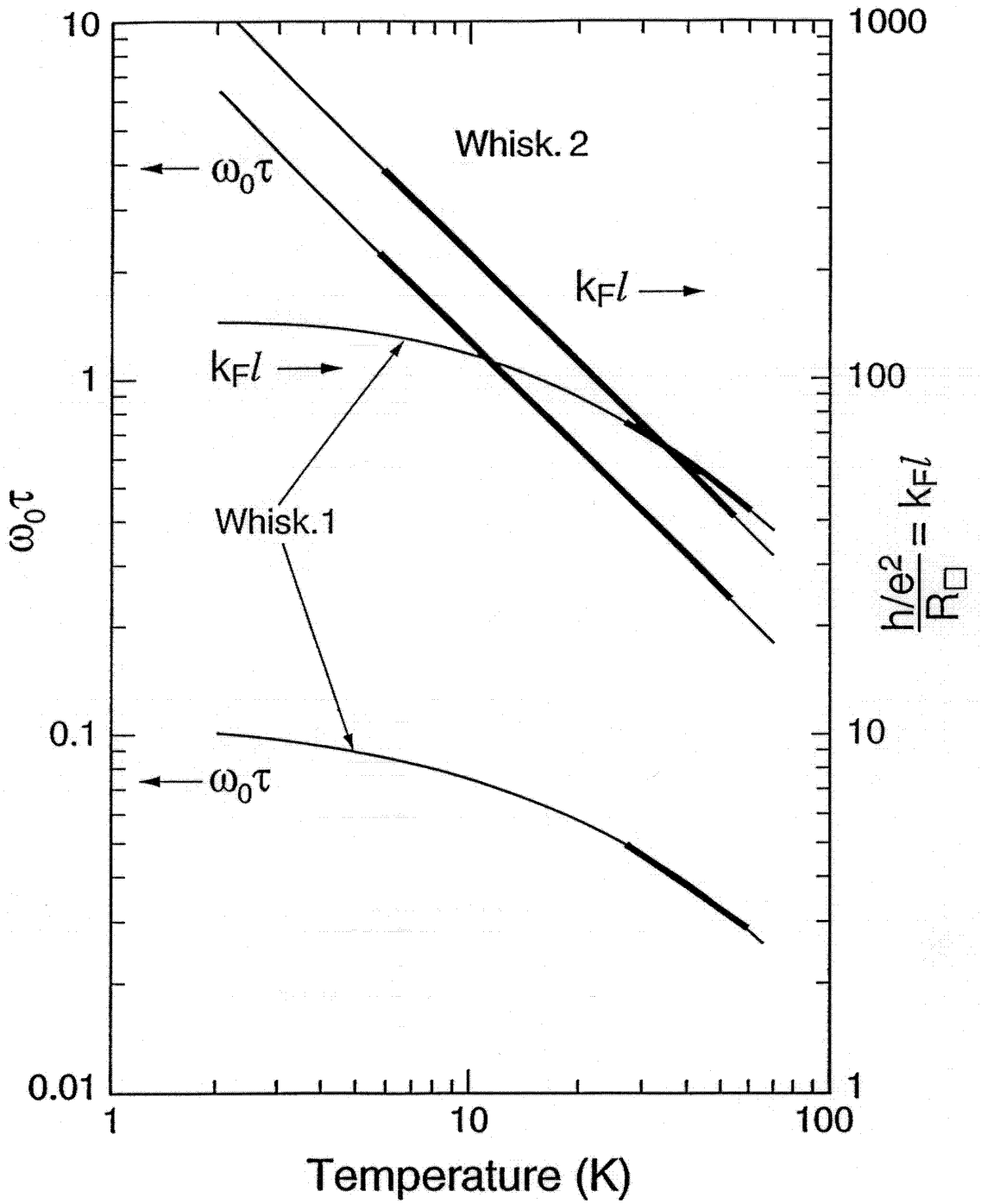
\end{document}